\documentclass[aps,pr,superscriptaddress,preprintnumbers,nofootinbib,10pt]{revtex4-2}

\usepackage{multirow}
\usepackage{amsmath}
\usepackage{amssymb}
\usepackage[dvipdf,dvips]{graphicx}
\usepackage{color}
\usepackage{hyperref}
\usepackage{url}
\usepackage{slashed}
\usepackage{subfigure}
\usepackage[usenames,dvipsnames]{xcolor}
\usepackage{amsmath}
\usepackage{amsfonts}
\usepackage{float} 
\usepackage{amssymb}
\usepackage{epsfig}
\usepackage{graphics}
\usepackage{euscript}
\usepackage{slashed}
\usepackage{epstopdf}
\usepackage[utf8]{inputenc}
\allowdisplaybreaks
\usepackage[normalem]{ulem}
\usepackage{pifont}
\usepackage{dsfont}
\usepackage{MnSymbol}
\usepackage{verbatim}
\usepackage{graphicx}
\usepackage{latexsym}
\usepackage{tikz-feynman}
\usepackage{bbold}
\newcommand{\cA}{\cal A}
\newcommand{\cB}{\cal B}
\newcommand{\cC}{\cal C}

\begin{document}

	\title{{\bf Bell's and Mermin's inequalities, entangled coherent states and unitary operators}}

	\author{S. P. Sorella} \email{silvio.sorella@gmail.com} \affiliation{UERJ $–$ Universidade do Estado do Rio de Janeiro,	Instituto de Física $–$ Departamento de Física Teórica $–$ Rua São Francisco Xavier 524, 20550-013, Maracanã, Rio de Janeiro, Brazil}

		\begin{abstract}
We elaborate on the recent proposal of employing unitary operators in Quantum Mechanics. The Bell and Mermin inequalities for entangled coherent states are scrutinized by making use of the unitary displacement operators. A violation of the Mermin inequality  close to the maximum allowed value is reported, in agreement with the existing literature. 
	\end{abstract}

	\maketitle

	\section{Introduction}\label{Intr} 
	
	In this work we pursue the investigation on the use of unitary operators in order to study the violation of both Bell-CHSH \cite{Bell64,CHSH69} and Mermin \cite{Mermin} inequalities. \\\\Unitary operators $\{ \cA \}$, ${\cA}^{\dagger} {\cA} = {\cA}{\cA}^{\dagger}=1$, are a special case of a more general class of operators, $\{ \cal Q \}$, known as normal operators:   ${\cal Q}^{\dagger} {\cal Q} = {\cal Q} {\cal Q}^{\dagger}$. Normal operators possess  a spectral representation \cite{Bratteli97}, a fundamental tool in Quantum Mechanics. As outlined in \cite{Erhard_2020,Hu_2017,Guimaraes:2024alk}, these operators are considered the natural candidates to generalize the notion of observable in Quantum Mechanics.  In particular, setting 
\begin{equation} 
{\cA} = {\cal M} + i {\cal N} \;, \qquad  {\cal M}={\cal M}^{\dagger} \;, \qquad {\cal N}={\cal N}^{\dagger} \;, \label{MN} 
\end{equation}
it follows that the unitary conditions yield 
\begin{equation} 
\left[ {\cal M}, {\cal N} \right] =0 \;, \qquad {\cal M}^2 + {\cal N}^2 =1 \;. \label{uc}
\end{equation}
Equation \eqref{uc} tells us that a unitary operator can be thought as being constituted by two Hermitian commuting operators $({\cal M},{\cal N})$ whose eigenvalues belong to the unit circle. \\\\Concerning now the Bell-CHSH and the Mermin inequalities, let us remind their  formulation
\begin{equation} 
\langle \psi_{AB} | \;{\cal C} | \;\psi_{AB} \rangle \;, \qquad {\cal C} = (A_1 + A_2) B_1 + (A_1-A_2)B_2 \;, \label{BAB}
\end{equation}	
and 
\begin{equation} 
\langle \psi_{ABC} | \;{\cal M}_3 | \;\psi_{ABC} \rangle \;, \qquad {\cal M}_3 = A_2 B_1C_1 + A _1B_2 C_1 + A_1B_1C_2- A_2B_2C_2 \;, \label{Mt}
\end{equation}	
where ${\cal M}_3$ stands for the Mermin operator of order three.  The state $\vert \psi_{AB} \rangle$ refers to a bipartite system $AB$, while $\vert \psi_{ABC} \rangle$ to a three-partite system $ABC$. The operators $(A_1,A_2)$, $(B_1,B_2)$, $(C_1,C_2)$ are Hermitian dichotomic operators fulfilling the conditions 
\begin{eqnarray} 
A_1 & =& A_1^{\dagger} \;, \quad A_2 = {A_2}^{\dagger} \;, \quad B_1  = B_1^{\dagger} \;, \quad B_2 = {B_2}^{\dagger} \;, \quad C_1 = C_1^{\dagger} \;, \quad C_2= {C_2}^{\dagger} \;, \nonumber \\[3mm]
A_1^2 & =& {A_2}^2 = B_1^2 = {B_2}^2=C_1^2 = {C_2}^{2} = 1 \;, \label{AB}
\end{eqnarray}
and
\begin{eqnarray}
\left[\; A_1,B_1\; \right] &= &  \left[\; A_2,B_1\; \right] =  \left[\; A_1,B_2\; \right] =   \left[\; A_2,B_2\; \right] =  \left[ \;A_1,C_1 \;\right] =  
\left[ \;A_1, C_2 \;\right] = 0 \;, \nonumber \\
 \left[ \; B_1,C_1 \; \right] & = & \left[ \; B_1, C_2 \; \right] = \left[ \; A_2,C_1 \; \right] = \left[ \; A_2, C_2 \; \right]= \left[ \; B_2,C_1 \; \right] = \left[ \; B_2, C_2 \; \right] = 0 \;,
 \nonumber \\[3mm]
\left[\; A_1,A_2\; \right] &\neq & 0\;, \quad \left[\; B_1,B_2\; \right] \neq  0 \;, \quad \left[\; C_1,C_2\; \right] \neq  0  \;. 
\label{ABBell}
\end{eqnarray}
One speaks of a violation of the Bell-CHSH inequality whenever 
\begin{equation} 
2 < \vert \langle \psi_{AB} | \;{\cal C} | \;\psi_{AB} \rangle \vert \le 2\sqrt{2} \;. \label{vb} 
\end{equation}
The value $2$ is the classical bound, while $2 \sqrt{2}$ is the maximum allowed violation, the so-called Tsirelson bound \cite{TSI}. In the case of the Mermin inequality of order three, eq.\eqref{Mt}, the violation occurs for  \cite{Mermin}
\begin{equation} 
2 < \vert \langle \psi_{ABC} | \;{\cal M}_3 | \;\psi_{ABC} \rangle \vert \le 4 \;. \label{vMI}
\end{equation} 
Making use of unitary operators means that, instead of employing operators as in eqs.\eqref{AB}.\eqref{ABBell}, computations are done with operators $({\cA}, {\cA}')$,  $({\cB}, {\cB}')$,  $({\cC}, {\cC}')$ subjects to the conditions 
\begin{eqnarray} 
	{\cA}^{\dagger}{\cA} & =  & 1 = {\cA}  {\cA}^{\dagger}   \;, \quad   {{\cA}'}^{\dagger} {{\cA}'} =  1 = {\cA}' {{\cA}'}^{\dagger}  \;, \nonumber \\[3mm]
	{\cB}^{\dagger} {\cB} & =  & 1 = {\cB} {\cB}^{\dagger}   \;, \quad   {{\cB}'}^{\dagger} {{\cB}'} =  1 = {\cB}' {{\cB}'}^{\dagger} \;, \nonumber \\[3mm]
	{\cC}^{\dagger} {\cC} & =  & 1 = {\cC} {\cC}^{\dagger}   \;, \quad   {{\cC}'}^{\dagger} {{\cC}'} =  1 = {\cC}' {{\cC}'}^{\dagger} \;, \nonumber \\[3mm]
	\quad \left[\; \cA, \cB\; \right] &= & 0\;, \quad \left[\; \cA', \cB\; \right] =  0 \;, \quad \left[\; \cA, \cB'\; \right] =  0 \;, \quad \left[\; \cA', \cB'\; \right] =  0 \;, \quad \left[\; \cA, \cC\; \right] =  0 \;, \quad \left[\; \cA, \cC'\; \right] =  0\nonumber \\[3mm]
		\quad \left[\;{ \cA}', \cC\; \right] &= & 0\;, \quad \left[\; \cA', \cC'  \; \right] =  0 \;, \quad \left[\; \cB, \cC\; \right] =  0 \;, \quad \left[\; \cB, \cC' \; \right] =  0 \;, \quad \left[\; \cB' , \cC\; \right] =  0 \;, \quad \left[\; \cB' , \cC' \; \right] =  0\nonumber \\[3mm]
	\left[\; \cA, \cA'\; \right] &\neq & 0\;, \quad \left[\; \cB, \cB'\; \right] \neq  0 \;,  \quad \left[\; \cC, \cC' \; \right] \neq  0 \;
	\label{ABBellu}
\end{eqnarray}
Replacing equations \eqref{AB}, \eqref{ABBell} by eqs.\eqref{ABBellu} requires a detailed re-examination of both classical and quantum bounds for the Bell-CHSH and Mermin inequalities. From the analysis performed in \cite{Guimaraes:2024alk}, it follows that the Bell-CHSH and Mermin inequalities can be consistently formulated in terms of unitary operators provided the reality conditions hold: 
\begin{eqnarray} 
\langle \psi_{AB} \vert \; {\cA}_i {\cB}_j \; \vert \psi_{AB}\rangle & = & {\rm real \; value} \quad \forall i,j=1,2  \;, \nonumber \\[3mm]
\langle \psi_{ABC} \vert \; {\cA}_i {\cB}_j  {\cC}_{k} \; \vert \psi_{ABC}\rangle & =&  {\rm real \; value} \quad \forall i,j,k =1,2  \;, \label{real}
\end{eqnarray}
where 
\begin{equation} 
\{ {\cA}_i, i=1,2 \} = ({\cA},{ \cA}' )  \;, \quad  \{ {\cB}_j, j=1,2 \} = ({\cB}, {\cB}' )  \;, \quad \{ {\cC}_k, k=1,2 \} = ({\cC}, {\cC} )  \;. \label{ccc}
\end{equation}
As we shall see, the conditions \eqref{real} are fulfilled in the case in which $(\vert \psi_{AB}\rangle, \vert \psi_{ABC}\rangle )$ are entangled coherent states and the operators $({\cA}, {\cA}')$, $({\cB}, {\cB}')$, $({\cC}, {\cC}')$ are identified with the unitary displacement operators.  This setup might result in new insights on the study of the properties of entangled coherent states for multipartite systems, while providing a useful application of the unitary operators. \\\\The work is organized as follows. In Sect.\eqref{BCH} we present the treatment of the Bell-CHSH via unitary displacement operators. Sect.\eqref{Mn} is devoted to the Mermin inequality of order three. Sect.\eqref{Conclusion} collects our conclusion.

\section{The Bell-CHSH inequality via unitary displacement operators}\label{BCH}

\begin{figure}[t!]
	\begin{minipage}[b]{0.6\linewidth}
		\includegraphics[width=\textwidth]{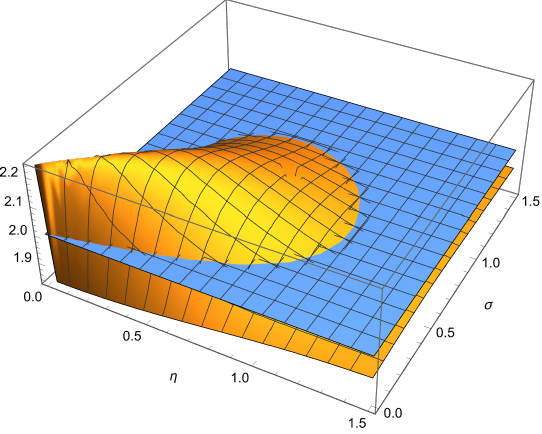}
	\end{minipage} \hfill
\caption{Behavior of the Bell-CHSH correlator $< \cal C>$ as a function of the parameters $(\eta, \sigma)$. The values of the complex numbers $(z,z',w,w')$ are as in equation \eqref{nv}. The region of the orange surface above the blue one corresponds to a violation of the Bell-CHSH inequality.}
	\label{Fig1}
	\end{figure}

The main tool of the present investigation is the use of the unitary displacement  operator 
\begin{equation} 
{\cal D}_a(z) = e^{z a^{\dagger} - z^* a} \;, \qquad {\cal D}_a(z){\cal D}^{\dagger}_a(z)={\cal D}^{\dagger}_a(z)  {\cal D}_a(z) =1 \;, \label{Da}
\end{equation}
where $z$ is a complex number, $z=x+iy$, and 
\begin{equation} 
[ a, a^{\dagger}] =1 \;, \qquad [a,a]=0 \;, \qquad [a^{\dagger}, a^{\dagger}] = 0 \;, \label{Aaa}
\end{equation}
stand for the criation and annihilation operators acting on the  Hilbert space ${\cal H}_A$ of the subsystem $A$. Similarly, the  displacement operator for the subsystem $B$ reads 
\begin{equation} 
{\cal D}_b(w) = e^{w b^{\dagger} - w^* b} \;, \qquad {\cal D}_b(w){\cal D}^{\dagger}_b(w)={\cal D}^{\dagger}_b(w)  {\cal D}_b(w) =1 \;, \label{Db}
\end{equation}
where $w=u+iv$ and $(b,b^{\dagger})$ 
\begin{equation} 
[ b, b^{\dagger}] =1 \;, \qquad [b,b]=0 \;, \qquad [b^{\dagger}, b^{\dagger}] = 0 \;,\qquad [a,b]=0 \;, \qquad [a, b^{\dagger}] =0 \;,  \label{ABab}
\end{equation}
being the ladder operators acting on the Hilbert space ${\cal H}_B$. The full Hilbert space is ${\cal H} = {\cal H}_A \otimes {\cal H}_B$. \\\\For further use, it is worth to write down the main relations which will be employed in the sequel: 
\begin{equation} 
{\cal D}_a(z) {\cal D}_a(i \eta) =.  e^{-i \eta \Re(z)}\; {\cal D}_a(z+ i\eta) \;, \label{eta}
\end{equation} 
where $\eta$ is a real number. Also 
\begin{eqnarray} 
{\cal D}^{\dagger}_a(i \eta) \;{\cal D}_a(z) \; {\cal D}_a(i \eta) & = &. e^{-2i \eta \Re(z)} \; {\cal D}_a(z) \;, \nonumber \\[3mm]
{\cal D}_a(i \eta) \;{\cal D}_a(z) \; {\cal D}_a(-i \eta) & = & e^{2i \eta \Re(z)} \; {\cal D}_a(z) \;, \nonumber \\[3mm]
{\cal D}_a(i \eta) \;{\cal D}_a(z) \; {\cal D}_a(i \eta) & =  & {\cal D}_a(z+ 2i \eta) \;. \label{rD}
\end{eqnarray}
Analogous equations hold for ${\cal D}_b(w)$. Let us proceed by specifying the entangled coherent state $\vert \psi_{AB} \rangle$ to  be used  in order to test the Bell-CHSH inequality: 
\begin{equation} 
\vert \psi_{AB} \rangle = {\cal N} \left( {\cal D}_a(i \eta) {\cal D}_b(i \sigma) - {\cal D}_a(-i \eta) {\cal D}_b(-i \sigma) \right) \vert 0 \rangle \;, \label{stab}
\end{equation} 
where $\sigma$ is a real parameter. For the normalization factor ${\cal N}$ one finds 
\begin{equation} 
{\cal N} = \frac{1}{\sqrt{2}} \frac{1}{\left( 1 - e^{-2(\eta^2 + \sigma^2)} \right)^{1/2}} \;. \label{nmz}
\end{equation} 
The next step is that os specifying the operators $(\cA, \cA')$, $(\cB,\cB')$. As already underlined, we shall make use of the unitary displacement operators, namely 
\begin{equation} 
{\cA} = {\cal D}_a(z)  \;, \qquad {\cA}'= {\cal D}_a(z') \;, \qquad {\cB} = {\cal D}_b(w) \;, \qquad {\cB}'= {\cal D}_b(w') \;. \label{opAB}
\end{equation} 
Therefore, for the Bell-CHSH inequality we have 
\begin{equation} 
\langle \psi_{AB} |\; {\cal C} \;| \psi_{AB} \rangle = \langle \psi_{AB}|\; \left( \left( {\cal D}_a(z) +  {\cal D}_a(z') \right)  {\cal D}_b(w) + \left( {\cal D}_a(z) -  {\cal D}_a(z') \right)  {\cal D}_b(w') \right) \; | \psi_{AB} \rangle  \;. \label{DBCH}
\end{equation}
We see that the displacement operators are employed for defining both the state $|\psi_{AB}\rangle$ and the Bell operators. This interesting possibility is consistently justified by noticing that the reality condition  \eqref{real} is fulfilled. In fact, a quick computation shows that 
\begin{equation}
\langle \psi_{AB} |\; {\cal D}_a(z) {\cal D}_b(w) \;| \psi_{AB} \rangle = {\cal N}^2 \left[  2 e^{-\frac{1}{2}(|z|^2 + |w|^2)} \cos\left( 2 \eta \Re(z) + 2 \sigma \Re(w) \right)  - e^{-\frac{1}{2} (|z+ 2i\eta|^2 + |w+ 2i\sigma|^2 ) }  - e^{-\frac{1}{2} (|z- 2i\eta|^2 + |w- 2i\sigma|^2 )   } \right]  \;. \label{rcond}
\end{equation}
Thus, for the Bell-CHSH correlator we get 
\begin{eqnarray} 
\langle \psi_{AB} |\; {\cal C} \;| \psi_{AB} \rangle & = & {\cal N}^2 \left[  2 e^{-\frac{1}{2}(|z|^2 + |w|^2)} \cos\left( 2 \eta \Re(z) + 2 \sigma \Re(w) \right)  - e^{-\frac{1}{2} (|z+ 2i\eta|^2 + |w+ 2i\sigma|^2 ) }  - e^{-\frac{1}{2} (|z- 2i\eta|^2 + |w- 2i\sigma|^2 )   } \right]  \nonumber \\
& + & {\cal N}^2 \left[  2 e^{-\frac{1}{2}(|z'|^2 + |w|^2)} \cos\left( 2 \eta \Re(z') + 2 \sigma \Re(w) \right)  - e^{-\frac{1}{2} (|z'+ 2i\eta|^2 + |w+ 2i\sigma|^2 ) }  - e^{-\frac{1}{2} (|z'- 2i\eta|^2 + |w- 2i\sigma|^2 )   } \right]  \nonumber \\
& + & {\cal N}^2 \left[  2 e^{-\frac{1}{2}(|z|^2 + |w'|^2)} \cos\left( 2 \eta \Re(z) + 2 \sigma \Re(w') \right)  - e^{-\frac{1}{2} (|z+ 2i\eta|^2 + |w'+ 2i\sigma|^2 ) }  - e^{-\frac{1}{2} (|z- 2i\eta|^2 + |w'- 2i\sigma|^2 )   } \right]  \nonumber \\
& - & {\cal N}^2 \left[  2 e^{-\frac{1}{2}(|z'|^2 + |w'|^2)} \cos\left( 2 \eta \Re(z') + 2 \sigma \Re(w') \right)  - e^{-\frac{1}{2} (|z'+ 2i\eta|^2 + |w'+ 2i\sigma|^2 ) }  - e^{-\frac{1}{2} (|z'- 2i\eta|^2 + |w'- 2i\sigma|^2 )   } \right]  \;. \label{fBCHS}
\end{eqnarray}
The meaning of the above expression is well captured by Fig.\eqref{Fig1}, where the value of the Bell-CHSH correlator is reported as a function of the parameters $(\eta, \sigma)$ which define the entangled coherent state $\vert \psi_{AB} \rangle$. The values of the complex numbers $z=x+iy$, $z'= x'+ iy'$, $w=u+iv$, $w'= u'+ iv'$ have been chosen after an optimization numerical analysis as: 
\begin{eqnarray} 
z & = & x+ iy \;, \qquad x= 0.01 \;, \qquad y= 0.12211 \nonumber \\
z'& = & x'+ iy'\;, \qquad x'= 0,01 \;, \qquad y'= -0.67795 \nonumber \\
w & = & u + i v \;, \qquad u = 0.001 \;, \qquad v = 0.122 \nonumber \\
w'& = & u'+ iv'\;, \qquad u'= 0.01 \;, \qquad v'=  -0.67826 \;. \label{nv}
\end{eqnarray}
The orange surface shows the behavior of $< {\cal C} >$, while the blue surface corresponds to the classical bound $< {\cal C}> =2$. One sees the existence of a rather big region in which the orange surface is above the blue one. All values of $(\eta, \sigma)$ belonging to this region correspond to violations of the Bell-CHSH inequality. The biggest value of the violation is of about $< {\cal C}>_{max}\approx 2.23$.

\section{The Mermin inequality}\label{Mn}
\begin{figure}[t!]
	\begin{minipage}[b]{0.6\linewidth}
		\includegraphics[width=\textwidth]{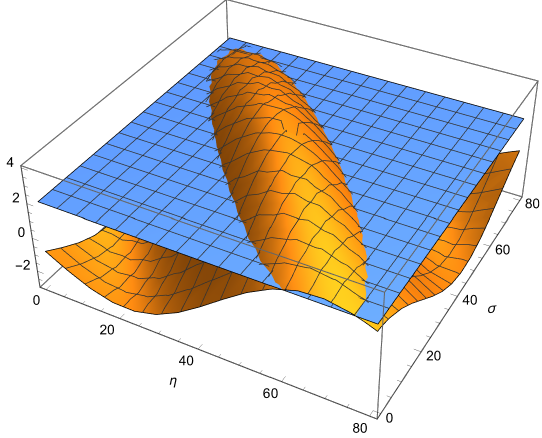}
	\end{minipage} \hfill
\caption{Behavior of the Mermin correlator $< {\cal M}_3>$ as a function of the parameters $(\eta, \sigma)$. The values of the complex numbers $(z,z',w,w',\zeta, \zeta', \tau)$ are as in equation \eqref{vm3}. The region of the orange surface above the blue one corresponds to a violation of the Mermin inequality.}
	\label{Fig2}
	\end{figure}

We move now  to the Mermin inequality. The Hilbert space is ${\cal H}= {\cal H}_A \otimes {\cal H}_B \otimes {\cal H}_C$. We need one more pair of ladder operators $(c,c^{\dagger})$: 
\begin{equation} 
{\cal D}_c(\zeta) = e^{\zeta c^{\dagger} - \zeta^* c} \;, \qquad {\cal D}_c(\zeta){\cal D}^{\dagger}_c(\zeta)={\cal D}^{\dagger}_c(\zeta)  {\cal D}_c(\zeta) =1 \;, \label{Dc}
\end{equation} 
with 
\begin{equation} 
[ c, c^{\dagger}] =1 \;, \qquad [c,c]=0 \;, \qquad [c^{\dagger}, c^{\dagger}] = 0 \;,\qquad [a,c]=0 \;, \qquad [a, c^{\dagger}] =0 \;,  \qquad [b,c]=0 \;, \qquad [b, c^{\dagger}] =0  \;. \label{CABabc}
\end{equation}
Here, we make use of the same class of entangled coherent states employed by \cite{JQ,DeFabritiis:2023bbs}, namely 
\begin{equation} 
\vert \psi_{ABC} \rangle = {\cal N}_{ABC} \left[ {\cal D}_a(i \eta) {\cal D}_b(i \sigma) {\cal D}_c(i\tau)- {\cal D}_a(-i \eta) {\cal D}_b(-i \sigma) {\cal D}_c(-i \tau) \right] \vert 0 \rangle  \;, \label{psm}
\end{equation} 
where 
\begin{equation} 
{\cal N}_{ABC} = \frac{1}{\sqrt{2}} \frac{1}{\left( 1 - e^{-2(\eta^2 + \sigma^2+ \tau^2)} \right)^{1/2}} \;. \label{nabc}
\end{equation} 
Also in this case, the reality condition \eqref{real} is fulfilled. In fact, it turns out that 
\begin{eqnarray} 
\langle \psi_{ABC} \vert \; {\cA} {\cB} {\cC} \; \vert \psi_{ABC} \rangle & =&  {\cal N}_{ABC}^2 \left[  2 e^{-\frac{1}{2}(|z|^2 + |w|^2+ |\zeta|^2)} \cos\left( 2 \eta \Re(z) + 2 \sigma \Re(w)  + 2\tau \Re(\zeta) \right)  \right] \nonumber \\
&-& {\cal N}_{ABC}^2 \left[ e^{-\frac{1}{2} (|z+ 2i\eta|^2 + |w+ 2i\sigma|^2 + |\zeta + 2 i \tau|^2) }  + e^{-\frac{1}{2} (|z- 2i\eta|^2 + |w- 2i\sigma|^2 + \zeta - 2 i \tau|^2)   } \right]  \;.  \label{rem3}
\end{eqnarray} 
Therefore, for the Mermin inequality one finds 
\begin{eqnarray} 
\langle \psi_{ABC} \vert \;{\cal M}_3 \; \vert \psi_{ABC} \rangle & =&  {\cal N}_{ABC}^2 \left[  2 e^{-\frac{1}{2}(|z'|^2 + |w|^2+ |\zeta|^2)} \cos\left( 2 \eta \Re(z') + 2 \sigma \Re(w)  + 2\tau \Re(\zeta) \right)  \right] \nonumber \\
&-& {\cal N}_{ABC}^2 \left[ e^{-\frac{1}{2} (|z'+ 2i\eta|^2 + |w+ 2i\sigma|^2 + |\zeta + 2 i \tau|^2) }  + e^{-\frac{1}{2} (|z'- 2i\eta|^2 + |w- 2i\sigma|^2 + |\zeta - 2 i \tau|^2)   } \right]  \nonumber \\
& + & 
 {\cal N}_{ABC}^2 \left[  2 e^{-\frac{1}{2}(|z|^2 + |w'|^2+ |\zeta|^2)} \cos\left( 2 \eta \Re(z) + 2 \sigma \Re(w')  + 2\tau \Re(\zeta) \right)  \right] \nonumber \\
&-& {\cal N}_{ABC}^2 \left[ e^{-\frac{1}{2} (|z+ 2i\eta|^2 + |w'+ 2i\sigma|^2 + |\zeta + 2 i \tau|^2) }  + e^{-\frac{1}{2} (|z- 2i\eta|^2 + |w'- 2i\sigma|^2 + |\zeta - 2 i \tau|^2)   } \right]  \nonumber \\
& + & 
 {\cal N}_{ABC}^2 \left[  2 e^{-\frac{1}{2}(|z|^2 + |w|^2+ |\zeta'|^2)} \cos\left( 2 \eta \Re(z) + 2 \sigma \Re(w)  + 2\tau \Re(\zeta') \right)  \right] \nonumber \\
&-& {\cal N}_{ABC}^2 \left[ e^{-\frac{1}{2} (|z+ 2i\eta|^2 + |w+ 2i\sigma|^2 + |\zeta' + 2 i \tau|^2) }  + e^{-\frac{1}{2} (|z- 2i\eta|^2 + |w- 2i\sigma|^2 + |\zeta' - 2 i \tau|^2)   } \right]  \nonumber \\
& - & 
 {\cal N}_{ABC}^2 \left[  2 e^{-\frac{1}{2}(|z'|^2 + |w'|^2+ |\zeta'|^2)} \cos\left( 2 \eta \Re(z') + 2 \sigma \Re(w')  + 2\tau \Re(\zeta') \right)  \right] \nonumber \\
&+& {\cal N}_{ABC}^2 \left[ e^{-\frac{1}{2} (|z'+ 2i\eta|^2 + |w'+ 2i\sigma|^2 + |\zeta' + 2 i \tau|^2) }  + e^{-\frac{1}{2} (|z'- 2i\eta|^2 + |w'- 2i\sigma|^2 + |\zeta' - 2 i \tau|^2)   } \right] \;.  \label{me3}
\end{eqnarray}
As done in the case of the Bell-CHSH we illustrate, in Fig.\eqref{Fig2} the behavior of the Mermin correlator $\langle M_3\rangle$ asa function of  $(\eta, \sigma)$, for the following values of the remaining parameters obtained through a maximization numerical analysis
\begin{eqnarray} 
z &=& 0.020091 - i \;0.00055757 \;, \qquad z'= 0.040244 - i \; 0.00114505 \;, \nonumber \\ 
w & = & 0.015207-i \;0.0000692535 \;, \qquad w'= 0.036766 -i \;0.00036440 \; \nonumber \\
\zeta & = & 0.0247087- i \;0.00050390 \;, \qquad \zeta'= 0.0437431 - i \;0.00087464 \;, \nonumber \\
\tau & = & 41.2201 \;. \label{vm3}
\end{eqnarray}
Again, the orange surface above the blue one  shows the region in parameter space in which the Mermin inequality is violated. It is interesting to observe that Fig.\eqref{Fig2} is in very good agreement with the results obtained by \cite{JQ}, where the Mermin inequality of order three for coherent states has been addressed by employing different operators $({\cA},{\cB}, {\cC})$. From Fig.\eqref{Fig2} one notices a region,  $\eta ->\rightarrow 38.8525$, $ \sigma \rightarrow  36.5831$, for which the violation of the Mermin inequality is very close to the maximum allowed value, namely: $\langle {\cal M}_3 \rangle_{max} = 3.99383 \approx 4$.  This is exactly the same pattern observed in  \cite{JQ}. 


\section{Conclusion}\label{Conclusion}

In this work we have pursued the study of the use of unitary operators in order to analyse both Bell-CHSH and Mermin inequalities. Such a possibility has been put forward by the authors \cite{Erhard_2020}, who have underlined the role of the normal operators as the natural class of operators which might be regarded as candidates to generalize the notion of observable in Quantum Mechanics. \\\\As far as unitary operators are concerned, we might give the following overview: 
\begin{itemize} 

\item as shown in \cite{Guimaraes:2024alk}, both Bell-CHSH and Mermin inequalities can be consistently formulated in terms of unitary operators, provided the reality conditions \eqref{real} are fulfilled. 

\item the conditions  \eqref{real} hold in the case of entangled coherent states, eqs.\eqref{stab},\eqref{psm}, when the operators $({\cA}, {\cB}, {\cC})$ are identified with the unitary displacement operators, eqs.\eqref{opAB}, \eqref{Dc}. This enables us to employ these operators for defining both states as well as Bell-CHSH and Mermin correlators.  

\item As one sees from equations \eqref{fBCHS},\eqref{me3}, the explicit calculations for both Bell-CHSH and Mermin inequalities are almost immediate, resulting in rather simple analytic expressions. 

\item From Fig.\eqref{Fig2} one learns that the violation of the Mermin inequality is nearly close to the maximum value of 4. This result is in agreement with the previous analysis worked out in \cite{JQ}. 

\item The displacement operators $({\cal D}_a(z), {\cal D}_b(w), {\cal D}_c(\zeta))$ are examples of Weyl operators, characterized by the Weyl algebra \cite{Bratteli97}, namely 
\begin{equation}
{\cal D}_a(z) \; {\cal D}_a(z') = e^{\frac{1}{2}(z {z'}^* - z^* z')} \; {\cal D}_a(z+z') \;. \label{weyla}
\end{equation}
As such, the present setup can be applied in the cases in which the Weyl algebra is present. As useful examples of such systems we may mention:\\\\{\it i)} the coordinate-momentum system $(q,p)$ in Quantum Mechanics \cite{pg}. For the Weyl operators, we have 
\begin{equation} 
{\cal U}_{\delta} = e^{i \delta q} \;, \qquad {\cal V}_\omega = e^{i \omega p} \;, \qquad [q,p]=i  \;. \label{UV}
\end{equation}
where $(\delta,\omega)$ are real parameters. \\\\{\it ii)} the Weyl operators of a relativistic Quantum Field Theory  \cite{DeFabritiis:2023tkh}: 
\begin{equation} 
{\cal A}_f = e^{i \int d^4x \;\varphi(x) f(x) } \;, \label{af}
\end{equation}
where $\varphi(x)$ is the scalar quantum field and $f(x)$ a smooth test function with compact support, see \cite{DeFabritiis:2023tkh} and refs. therein. These examples give an idea of the large applicability of the present framework. 

\end{itemize}

\section*{Acknowledgments}
	The author would like to thank P. De Fabritiis, F. Guedes, M.S. Guimaraes and I. Roditi for discussions and collaborations. The Brazilian agencies CNPq and FAPERJ  are gratefully acknowledged  for financial support.  S.P.~Sorella is a CNPq researcher under contract 301030/2019-7.

\newpage

\vspace{4cm}


\end{document}